\documentclass[pra,twocolumn,superscriptaddress,showpacs,amsmath,amssymb,floatfix]{revtex4}
\usepackage{bbm}

\usepackage{graphicx}%
\usepackage{dcolumn}
\usepackage{amsmath}
\usepackage{bm}
\usepackage{color}
\usepackage{subeqnarray}
\newcommand{\be}{\begin{equation}}
\newcommand{\ee}{\end{equation}}
\newcommand{\bey}{\begin{eqnarray}}
\newcommand{\eey}{\end{eqnarray}}
\newcommand{\bw}{\begin{widetext}}
\newcommand{\ew}{\end{widetext}}

\newcommand{\ra}{\rangle}
\newcommand{\la}{\langle}

\newcommand{\pp}{\partial}

\newcommand{\ba}{\begin{array}}
\newcommand{\ea}{\end{array}}
\newcommand{\bi}{\begin{itemize}}
\newcommand{\ei}{\end{itemize}}
\newcommand{\bem}{\begin{enumerate}}
\newcommand{\eem}{\end{enumerate}}
\newcommand{\im}{{\rm i}}

\begin{document}

 \title {Scaling behavior for a class of quantum phase transitions}

\author{Wen-ge Wang}
\email{wgwang@ustc.edu.cn}
\affiliation{Department of Modern Physics, University of Science and Technology
of China, Hefei 230026, China}
\author{Pinquan Qin}
\affiliation{Department of Modern Physics, University of Science and Technology
of China, Hefei 230026, China}
\author{Qian Wang}
\affiliation{Department of Modern Physics, University of Science and Technology
of China, Hefei 230026, China}
\author{Giuliano Benenti}
\affiliation{CNISM $\&$ Center for Nonlinear and Complex Systems,
Universit\`a degli Studi dell'Insubria, Via Valleggio 11, 22100 Como, Italy}
\affiliation{Istituto Nazionale di Fisica Nucleare, Sezione di Milano,
via Celoria 16, 20133 Milano, Italy}
\author{Giulio Casati}
\affiliation{CNISM $\&$ Center for Nonlinear and Complex Systems,
Universit\`a degli Studi dell'Insubria, Via Valleggio 11, 22100 Como, Italy}
\affiliation{Istituto Nazionale di Fisica Nucleare, Sezione di Milano,
via Celoria 16, 20133 Milano, Italy}

\date{\today}

 \begin{abstract}
We show that for quantum phase transitions with a single bosonic
zero mode at the critical point, like the Dicke model and the
Lipkin-Meshkov-Glick model,
metric quantities such as fidelity,
that is, the overlap between two ground states corresponding to two values
$\lambda_1$ and $\lambda_2$ of the controlling parameter $\lambda$,
only depend on the ratio
$\eta=(\lambda_1-\lambda_c)/(\lambda_2-\lambda_c)$, where
$\lambda=\lambda_c$ at the critical point.
Such scaling property is valid also for time-dependent quantities
such as the Loschmidt echo,
provided time is measured in units of the inverse frequency of the critical
mode.
 \end{abstract}
 \pacs{64.70.Tg; 05.70.Jk; 03.67.-a; 32.80.Qk}


 \maketitle

\section{Introduction}
\label{sec:intro}

 Due to the quantum nature, a Quantum Phase Transition (QPT)~\cite{sachdev},
defined as a drastic change of fundamental
 properties of ground states, may have properties quite different from thermal phase transitions.
 This point has received much attention in recent years.
In particular,
some concepts and quantities in the field of
quantum information, e.g., entanglement and fidelity,
 have been found quite useful in characterizing the occurrence of a QPT.
For example, the overlap between two ground states corresponding to
two nearby values $\lambda_1$ and $\lambda_2$ of the controlling
parameter $\lambda$,
\be
L_p (\lambda_1,\lambda_2) = |\langle 0_{\lambda_1}|0_{\lambda_2} \rangle |,
\label{lpe}
\ee
has been proposed as a probe of quantum criticality~\cite{zan2006}:
the dramatic change of the wave function at a QPT implies a decrease
of the overlap $L_p$ in the neighborhood of the critical point.
Hence, the fidelity $L_p$ can be used to
detect the occurrence of a QPT (see
\cite{gu2010,zan2006,zho2008,zan2007,coz2007,cozz2007,buo2007,zana2007,RD11}
and references therein).

 Similarly to thermal phase transitions, an important aspect
 of QPTs is the
 dependence of relevant quantities on the controlling parameter $\lambda$.
 For example, the characteristic energy scale
 usually takes the form $|\lambda - \lambda_c|^{\varphi}$, with
$\lambda_c$  indicating the critical point and
$\varphi > 0$ a critical exponent.
This is for the case in which only one value of the
parameter $\lambda$ is of relevance.
For quantities like
fidelity, which instead depend on two
values $\lambda_1$ and $\lambda_2$ of
the controlling parameter, one should understand
whether their behavior in the critical region
encodes universal properties about a QPT.
With regard to the fidelity, the question is whether its
drop near a QPT can be used not only to detect the QPT
itself but also to determine the critical
exponents~\cite{RD11}.
In this context, understanding scaling properties
is relevant.

When studying fidelity, either as the overlap
(\ref{lpe}) of ground states or as the survival probability of an
initial state prepared in the ground state $|0_{\lambda_2}\rangle$ of
Hamiltonian $\hat{H}(\lambda_2)$ and evolved under
a different Hamiltonian $\hat{H}(\lambda_1)$,
two values $\lambda_1$ and $\lambda_2$
of the controlling parameter are involved.
For the sake of clearness, in what follows, we use the name fidelity
for the former quantity, $L_p(\lambda_1,\lambda_2)$, and use
Loschmidt Echo (LE) for the latter one~\cite{footnote}.
 An interesting question is about the dependence of fidelity and
LE on $\lambda_1$
 and $\lambda_2$.
 In the case that fidelity $L_p(\lambda_1,\lambda_2)$ goes to
 zero when $\lambda_1$ approaches $\lambda_c$ for a fixed $\lambda_2$,
 it is clear that for each $\lambda_2'$ there exists a value $\lambda_1'$
 such that $L_p(\lambda_1,\lambda_2)=L_p(\lambda_1',\lambda_2')$.
 This implies that the relative positions of
 $\lambda_1$ and $\lambda_2$ with respect to $\lambda_c$,
 rather than their exact positions, play the crucial role.
 Hence, the question arises of whether fidelity may be invariant
 under rescaling of the controlling parameter.

In this paper, we show that for QPTs
possessing only one bosonic zero mode at the critical point,
 metric quantities like fidelity only depend on the ratio
$\eta = (\lambda_1-\lambda_c)/(\lambda_2-\lambda_c)$.
 That is, these physical quantities are invariant under
 linear rescaling of the controlling parameter with respect to
 the critical point.
We also show that such scaling property is valid 
for time-dependent quantities such as the LE,
provided time is measured in units of the inverse frequency of the critical
mode.
The class of QPTs possessing such features includes important physical models, like
the Dicke~\cite{Emary03} and
the Lipkin-Meshkov-Glick (LMG) model~\cite{LMG}.

The article is organized as follows. 
In Sec.~\ref{sec:scaling} we discuss our scaling argument
for static metric quantities like fidelity, extending 
in Sec.~\ref{sec:timedependent} such
scaling to time-depending quantities like the LE.
The scaling for time-dependent quantities is then 
illustrated by means of the semicassical theory. 
We then illustrate the fidelity and LE 
scaling in two relevant physical models, 
the Dicke model (Sec.~\ref{sec:Dicke}) and the LMG
model (Sec.~\ref{sec:LMG}).
We finish with concluding remarks in Sec.~\ref{sec:discussion}.

\section{Scaling argument}
\label{sec:scaling}

As only the lowest energy levels are concerned close to the critical
point, we assume that the Hamiltonian describing a QPT
can be approximately written
in terms of $n$ harmonic oscillators:
\begin{equation}\label{Hhar}
\hat{H}(\lambda) = \sum_{i=1}^n e_{i}(\lambda)
\hat{c}_{i}^{\dag}(\lambda) \hat{c}_{i}(\lambda),
\end{equation}
where $\hat{c}_i^{\dag}(\lambda)$ and $\hat{c}_i(\lambda)$ are
\emph{bosonic} creation
and annihilation operators for the $i$-th mode.
The ground state $|0_\lambda\ra$ for the parameter $\lambda$
 is defined by $\hat{c}_{i}(\lambda)|0_\lambda\ra =0$.
 For two parameter values $\lambda_1$ and $\lambda_2$, one may write
\begin{equation}\label{c-sum}
\hat{c}_i^\dag(\lambda_1) =  \sum_{j=1}^n \left[P_{ij}
\hat{c}_j^\dag (\lambda_2)
+ Q_{ij} \hat{c}_j(\lambda_2)\right],
\end{equation}
 where $P_{ij}$ and $Q_{ij}$ are functions of $\lambda_1$ and $\lambda_2$,
 with $P_{ij} = \delta_{ij}$ and $Q_{ij}=0$ for $\lambda_1 = \lambda_2$.
We discuss the case in which there is only one zero mode
at the critical point:
$e_1(\lambda) \sim |\lambda - \lambda_c|^{\varphi}$ for $\lambda$
close to $\lambda_c$, while $e_i(\lambda_c)\ne 0$ for $i\ne 1$.
 In this case, Eq.~(\ref{c-sum}) reduces to:
\begin{equation}\label{cc1}
\hat{c}_1^\dag(\lambda_1) = P_{11} \hat{c}_1^\dag (\lambda_2) +
Q_{11} \hat{c}_1(\lambda_2),
\end{equation}
with a corresponding expression for $\hat{c}_1(\lambda_1)$.
 From the bosonic commutation relations it follows
that $|P_{11}|^2 - |Q_{11}|^2=1$.
 Let us write explicitly the phases of $P_{11}$ and $Q_{11}$,
 as $P_{11}=|P_{11}| {e}^{\im (\theta_c+\theta_r)}$,
 $Q_{11}=|Q_{11}| {e}^{\im (\theta_c-\theta_r)}$.
 In the representation of $\hat{H}(\lambda_2)$,
 the change of the pair $(\hat{c}_1 (\lambda_2), \hat{c}_1^\dag (\lambda_2))$ to
 $(e^{-\im \theta_r}\hat{c}_1 (\lambda_2), e^{\im \theta_r}\hat{c}_1^\dag (\lambda_2))$
 does not bring any change to the physics,
 hence, the phase $\theta_r$ can be
 absorbed by $\hat{c}_1 (\lambda_2)$ and $\hat{c}_1^\dag (\lambda_2)$.
 The phase $\theta_c$ is the relative phase between the
 pair of operators $(\hat{c}_1 (\lambda_2), \hat{c}_1^\dag (\lambda_2))$ at $\lambda_2$ and
 the pair $(\hat{c}_1 (\lambda_1), \hat{c}_1^\dag (\lambda_1))$ at $\lambda_1$,
 which generates relative phases between the set of basis states
 $|n_{\lambda_1} \ra =(\hat{c}_1^\dagger(\lambda_1))^n |0_{\lambda_1}\ra|/\sqrt{n!}$
 and
 $|m_{\lambda_2} \ra =(\hat{c}_1^\dagger(\lambda_2))^m |0_{\lambda_2}\ra|/\sqrt{m!}$.

Let us consider a physical quantity $A$
depending on two values $\lambda_1$ and $\lambda_2$ of the controlling parameter
(for instance, $A$ might be the fidelity),
written in the vicinity of the critical point
as a function of the annihilation
operators for the zero mode:
\be
A=\langle 0_{\lambda_i} |
\hat{A}(\hat{c}_{1}(\lambda_1), \hat{c}_{1}(\lambda_2))
|0_{\lambda_j}\rangle, \;\;(i,j=1,2).
\label{A}
\ee
 In what follows, we focus on quantities $A$ that do not
 depend on the phase $\theta_c$ \cite{foot-neta}.
 In particular, \emph{metric quantities},
 like $C_{nm}\equiv|\la n_{\lambda_1}|m_{\lambda_2}\ra|$,
 belong to this class.
Such quantities include, for instance, the fidelity $L_p=C_{00}$ and the
participation ratio $\chi$ of an eigenstate of e.g. $\hat{H}({\lambda_2})$,
$|m_{\lambda_2}\rangle=\sum_{n} \la n_{\lambda_1} | m_{\lambda_2}\ra
|n_{\lambda_1}\rangle$,
with respect to the basis of the eigenstates
of $\hat{H}(\lambda_1)$; by definition,
$\chi=1/\sum_{n} |\la n_{\lambda_1} | m_{\lambda_2}\ra|^4=
1/\sum_{n} C_{mn}^4$.
In the study of these quantities, we can take $\theta_c=0$.
Then, since the phase $\theta_r$ can be
 absorbed by $\hat{c}_1 (\lambda_2)$ and $\hat{c}_1^\dag (\lambda_2)$,
$P_{11}$ and $Q_{11}$ are just their absolute values, with $P_{11}\ge 1$.
Using the ground state definition
$\hat{c}_1(\lambda_1)|0_{\lambda_1}\rangle=0$,
Eq.~(\ref{cc1}),
and the expansion
$|0_{\lambda_1}\rangle=\sum_m\langle m_{\lambda_2} |
0_{\lambda_1}\rangle| m_{\lambda_2}\rangle$, we
can express $|0_{\lambda_1}\rangle$ as a function of
$P_{11}$, $Q_{11}$, $\hat{c}_1^\dagger(\lambda_2)$,
and $|0_{\lambda_2}\rangle$.
After inserting the obtained expression for $|0_{\lambda_1}\rangle$
into (\ref{A}), we find that $A$ is a function of
$P_{11}$ and $Q_{11}$ only.

The dependence of $P_{11}$
on $\lambda_1$ and $\lambda_2$ can be written as $P_{11}=F(\Delta \lambda_1, \Delta \lambda_2)$,
 where $\Delta \lambda_i = \lambda_i -\lambda_c$ ($i=1,2$).
We study $\lambda_1$ and $\lambda_2$ belonging to the same phase, so
that $\eta=\Delta\lambda_1/\Delta\lambda_2>0$.
We assume, as it is natural for a QPT with an infinitely-degenerate zero mode
at the critical point, that
 $F(\Delta \lambda_1, \Delta \lambda_2)$ goes to infinity in the limit
 $\Delta \lambda_1 \to 0$ with $\Delta \lambda_2 \ne 0$, as well as in the limit
 $\Delta \lambda_2 \to 0$ with $\Delta \lambda_1 \ne 0$.
For the sake of simplicity,
we assume that $F$ is a monotonic function of $\Delta \lambda_2$,
 when $\Delta \lambda_2$ changes from a given $\Delta \lambda_1$
to 0~\cite{foot-function relation}.
Then, given $F(\Delta \lambda_1, \Delta \lambda_2)=d$, for any $\lambda_1'$
 there must exist a $\lambda_2'$ such that
$F(\Delta \lambda_1', \Delta \lambda_2')=d$.
 This implies that there exists a function
 $\Delta \lambda_2 = g(\Delta \lambda_1,d)$,
 such that $F(\Delta \lambda_1 ,g(\Delta \lambda_1,d))= d$
for any $\Delta \lambda_1$, hence,
 \begin{equation}\label{pF}
 \partial F / \partial \Delta \lambda_1 + (\partial F /
\partial \Delta \lambda_2)
 (\partial g / \partial \Delta \lambda_1) =0.
 \end{equation}
 For a given $d$ and a sufficiently small $\Delta \lambda_1$,
 Taylor expansion reads $g= g_0 + g' \Delta \lambda_1$, where
$g' = \partial g / \partial \Delta \lambda_1$
 with $d$ fixed.
 We recall that in the limit $\Delta \lambda_1 \to 0$,
 $F(\Delta \lambda_1, \Delta \lambda_2)$ goes to infinity
 if $\Delta \lambda_2$ is non-zero.
 Hence, for any given $d$, $\Delta \lambda_2$ must go to zero in order that
$\lim_{\Delta\lambda_1\to 0}F(\Delta\lambda_1,\Delta\lambda_2)=d$.
 That is, $\lim_{\Delta \lambda_1 \to 0} g(\Delta \lambda_1,d) =0$,
 as a result, $g_0=0$.
 Therefore, when the above Taylor expansion works for all fixed values of
 $d$, we have $ \partial g / \partial \Delta \lambda_1 =
g/\Delta \lambda_1
 = \Delta \lambda_2 /\Delta \lambda_1$.
 Substituting this result into Eq.~(\ref{pF}), we find
 \begin{equation}\label{}
 \Delta \lambda_1 \! \ \partial F / \partial \Delta \lambda_1 +
 \Delta \lambda_2 \! \ \partial F / \partial \Delta \lambda_2  =0.
 \end{equation}
 This equation has the solution $F= F(\ln \Delta \lambda_1 -
\ln \Delta \lambda_2)
 = F(\ln \eta)$. Hence, $P_{11}$ is a function of $\eta$.
 Then, due to the relation $P_{11}^2-Q_{11}^2=1$, $Q_{11}$ is
also a function of $\eta$.
Since the quantity $A$ is a function of $P_{11}$ and $Q_{11}$, we can conclude
that $A$ only depends on the ratio
$\eta=(\lambda_1-\lambda_c)/(\lambda_2-\lambda_c)$

\section{Time-dependent quantities}
\label{sec:timedependent}

We now consider time-dependent metric quantities $A(t)$,
with the dynamics described by Hamiltonian
(\ref{Hhar}).
Since the  frequencies
$\omega_i(\lambda)=e_i(\lambda)/\hbar$ depend on
 $\lambda$,
 $A(t)$ usually cannot be a function of $\eta$ only.
However, for systems with a single zero mode at the critical
point, the $\eta$-scaling  still applies,
provided time is rescaled: $t\to\tau \equiv \omega_1(\lambda)t$.

As an illustration we show in the following that in the vicinity of a QPT with a single zero
mode at the critical point the decay of the quantum Loschmidt echo
depends only on the scaling parameter $\eta$ and the rescaled time $\tau$.
The LE gives a measure for the stability of the quantum motion
under slight variation of the Hamiltonian
\cite{peres84,nie2000,ben2004}. It is defined
by $M_L(t) = |m(t)|^2 $, where
\be m(t) = \la \Psi_0|{\rm exp}(i \hat{H}(\lambda_2) t/ \hbar )
{\rm exp}(-i\hat{H}(\lambda_1)t / \hbar)
 |\Psi_0 \ra . \label{mat} \ee
Here, $\hat{H}(\lambda_1)=
\hat{H}(\lambda_2)+\epsilon \hat{V}$, with $\epsilon=\lambda_1-\lambda_2$.
Extensive investigations have been performed in recent years,
to understand the decaying behavior
of the LE in different regimes, depending
on the chaotic or integrable nature of the dynamics, on the
system's dimensionality, and on the perturbation strength
(see Refs.\cite{jal2001,jac2001,jac2002,cuc2002,STB03,wan2005,cer2002,
VH03,pro2002,wan2004,wang2005,wan2007,BC02,Gorin-rep,jacquodreport,Wang10}
and references therein).
Furthermore recent investigations have
shown that  the LE
may be employed to characterize QPTs,
since it exhibits extra-fast decay in the vicinity of
critical points \cite{Quan06,LE-qpt1,LE-qpt2,Rossini07,Peng08,Wang10}.

Here we consider a system initially prepared in the
ground state $|0_{\lambda_2}\ra $
of $\hat{H}(\lambda_2)$. Then the LE is in fact the survival probability,
\be M_L(t) =|\la 0_{\lambda_2} |e^{-i\hat{H}(\lambda_1)t/\hbar}|
0_{\lambda_2}\ra |^2. \ee
In the critical region, $\hat{H}(\lambda_2)$ represents a harmonic
oscillator and, therefore, its ground state can be written as a
Gaussian wave packet.
As shown in Ref.~\cite{wan2007}, when the classical motion
is periodic with a period $T_p$,
semiclassical theory predicts that
for $t>T_p$ the LE has an initial Gaussian
decay followed by a power law decay.
Indeed, to a second-order term of perturbation expansion,
\be M_L(t) \simeq {b_0}{(1+\xi^2 t^2)^{-1/2}}
e^{ -\Gamma t^2 /(1+\xi^2 t^2)} , \label{Mt-sc} \ee
where $b_0\sim 1$,
$\Gamma = ( \frac{\epsilon  W}{\hbar} \frac{\pp U}{\pp p_0}  )^2/2$,
$\xi = | \frac{\epsilon W^2}{2\hbar} \frac{ \pp^2 U} { \pp p_0^2}|$,
with the derivatives evaluated at the center $p_0$ of the
initial Gaussian wave packet.
Here, $U= \frac 1{T_p} \int_0^{T_p} V\,dt$ and
$W $ is a measure of the width of the initial Gaussian
packet in the momentum space.
It is seen that $M_L$ has a Gaussian decay
$e^{-\Gamma t^2}$ for short times and
a $1/{\xi t}$ decay for long times.

Let us consider the case in which $\hat{H}$ has, close to the
critical point, two lowest energy relevant modes, of which
only the first one has zero frequency at the critical point.
The first mode moves very slowly, so that effectively the
system moves periodically with the frequency of the second mode.
That is, for times much shorter than the period $T_1$
of the first mode (which diverges at the critical point),
the classical motion
is approximately periodic with the period of the second mode.
Hence, the period
$T_p = 2\pi/ \omega_2(\lambda_1)$.
Note that for long times the first mode dominates and the
LE oscillates with a period related to $T_1 = 2\pi / \omega_1(\lambda_1)$ (detailed later).
Therefore, the above semiclassical prediction works within times
longer than $T_p$ and shorter than $T_1$.

Starting from the classical expression of the Hamiltonian,
\be H(\lambda) = \omega_1(\lambda) I_1(\lambda) +
\omega_2(\lambda) I_2(\lambda).
\ee
we obtain
\begin{eqnarray} \label{}
 \xi t = \frac{\xi}{\omega_1(\lambda_1)} \tau
\sim \frac{\partial^2}{\partial p_0^2}
\la \frac{H(\lambda_1)-H(\lambda_2)}{e_1(\lambda_1)} \ra \tau
 \\  \simeq \frac{\partial^2}{\partial p_0^2}
\left \la \frac{I_1(\lambda_1)}{\hbar}
- \frac{e_1(\lambda_2)}{e_1(\lambda_1)}
\frac{I_1(\lambda_2)}{\hbar} +
\frac{\Delta H_2}{e_1(\lambda_1)} \right \ra \tau,
\end{eqnarray}
where $\Delta H_2 = I_2(\lambda_1)\omega_2(\lambda_1) -
I_2(\lambda_2)\omega_2(\lambda_2)$.
The second mode has no singularity at $\lambda_c$, hence
from Taylor expansion of $\Delta H_2$ we
obtain $\Delta H_2 \sim (\lambda_1 - \lambda_2)$.
Hence, $\frac{\Delta H_2}{e_1(\lambda_1)} \sim |1-\eta^{-1}|^\varphi
|\lambda_1 -\lambda_2|^{1-\varphi}$.
Therefore, when $\lambda_1$ is sufficiently close to $\lambda_2$,
$\frac{\Delta H_2}{e_1(\lambda_1)}$ can be neglected for $\varphi <1$ and $\eta \ne 0$.
Then, since $I_1(\lambda_1)$ and $I_1(\lambda_2)$ have no
singularity at $\lambda_c$ and $\frac{e_1(\lambda_2)}{e_1(\lambda_1)} =
\eta^{-\varphi} $,
we find that in the very neighborhood of $\lambda_c$,
$\xi t \simeq F(\eta ) \tau$.
Similarly, $\Gamma t^2 $ can be written as $G(\eta) \tau^2$.
We can therefore conclude that
\be M_L(t) \simeq {b_0}{(1+(F(\eta) \tau)^2)^{-1/2}}
e^{ -G(\eta)\tau^2 /(1+(F(\eta) \tau)^2)} \label{Mt-sc-sa} \ee
is a function of $\eta$ and of the rescaled time $\tau$.

\section{Scaling for the Dicke model}
\label{sec:Dicke}

This model~\cite{Emary03}
provides a physically significant example of our scaling behavior.
It describes the interaction between a single bosonic mode
and a collection of $N$ two-level atoms and finds applications
in quantum optics, condensed matter physics and quantum
information.
In terms of the collective operator ${\bf \hat{J}}$ for the $N$ atoms,
the Dicke Hamiltonian is written as (hereafter we take $\hbar =1$)
 \be
 \hat{H}(\lambda )=\omega_{0}\hat{J}_{z}
+\omega \hat{a}^{\dag}\hat{a}
+ ({\lambda}/{\sqrt{N}})(\hat{a}^{\dag}+\hat{a})(\hat{J}_{+}+\hat{J}_{-}).
\label{DH} \ee

In the thermodynamic limit $N\to \infty$, the system undergoes a QPT at
$\lambda_c = \frac 12 \sqrt{\omega \omega_0}$,
with a normal phase for $\lambda <\lambda_c$
and a super-radiant phase for $\lambda > \lambda_c$.
The Hamiltonian can be diagonalized in this limit
\cite{Emary03}, taking, up to a constant energy term, the form
(\ref{Hhar}), with $n=2$ modes. In the normal phase, the energies of the two
harmonic oscillators read
 \be
 e_{1,2}({\lambda}) = \Big\{\frac{1}{2}[(\omega^{2} +
\omega^{2}_{0}) \pm \sqrt{(\omega^{2}_{0}-\omega^{2})^{2}
  + 16\lambda^{2}\omega\omega_{0}}]\Big\}^{1/2},
  \label{ee}
 \ee
ordered so that $e_1 (\lambda)< e_2(\lambda)$.
It is seen that $e_{1}({\lambda})=0$ for $\lambda= \lambda_c$,
hence, the ground level of $\hat{H}(\lambda_c)$ is infinitely degenerate
and the system undergoes a QPT at $\lambda_c$.
On the other hand, $e_2(\lambda)\ne 0$ at the critical point.
In the super-radiant phase, the Dicke Hamiltonian can still be
diagonalized in the thermodynamic limit, resulting in a
two-mode form, with the energies
 \be
 e_{1,2}(\lambda) = \Big\{\frac{1}{2}[\omega^{2} + \frac{\omega^{2}_{0}}{\mu^2}
 \pm \sqrt{\big(\frac{\omega^{2}_{0}}{\mu^2}-\omega^{2}\big)^{2}
  + 4\omega^2\omega^2_{0}}]\Big\}^{1/2},
 \label{ep}
 \ee
where $\mu\equiv \omega \omega_0/4\lambda^2$ and
$e_{1}(\lambda)<e_{2}(\lambda)$.
It is easy to see that $e_{1}(\lambda)=0$ and
$e_{2}(\lambda)\ne 0$ for $\lambda= \lambda_c$.
Thus, the ground level of $\hat{H}(\lambda_c)$ is also infinitely
degenerate (and with a single zero mode) at the critical point
from the super-radiant-phase side.

\begin{figure}
\includegraphics[width=\columnwidth]{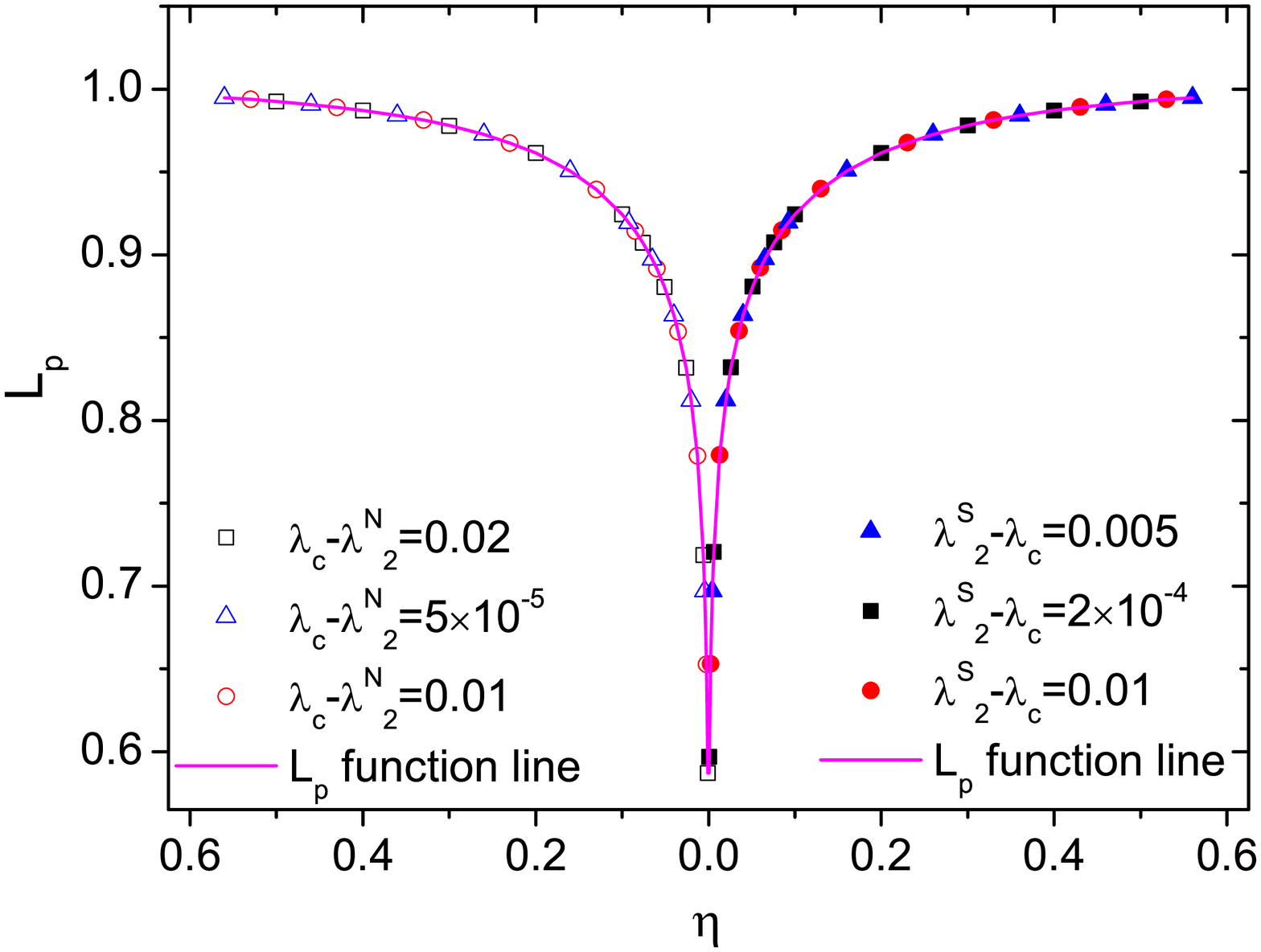}
\caption{(Color online)
Dependence of the fidelity $L_p$ on the scaling parameter
$\eta$ for different values of $\lambda_1$ and $\lambda_2$.
Note that the values of $\eta$ on both sides of the zero point
are positive.
Symbols on the left hand side of $\eta=0$ represent fidelity
in the normal phase, with the superscript $N$ of $\lambda_2$ standing
for normal phase
(open symbols).
Symbols on the right hand side of $\eta=0$ are for the
super-radiant phase (superscript $S$ of $\lambda_2$, full symbols).
Data are in agreement
with the analytical result of Eq.~(\ref{lpf}) (solid curve).
}
 \label{lp}
 \end{figure}

As can be shown both analytically and numerically the fidelity in this model
uniquely depends on the scaling parameter $\eta$. Indeed,
in both phases, it has been found 
(see Aappendix~\ref{sec:disckeanalytical}) that
 \be
L_p
=\frac{\sqrt{2}\sqrt[4]{\eta^\varphi}}{\sqrt{\eta^\varphi+1}}
= \frac{\sqrt{2}\sqrt[8]{\eta}}{\sqrt{\sqrt{\eta}+1}},
 \label{lpf}
 \ee
with the critical exponent $\varphi =1/2$
($\omega_1(\lambda)\sim |\lambda-\lambda_c|^{1/2}$)~\cite{note-zanardi}.
For $\eta\ll 1$, $L_p\propto \eta^{\varphi/4}$.
The analytical result (\ref{lpf}) is
in agreement with numerical simulations shown in Fig.~\ref{lp}:
data for different values of $\lambda_1$ and $\lambda_2$ collapse
on a single universal curve.

\begin{figure}
 \includegraphics[width=\columnwidth]{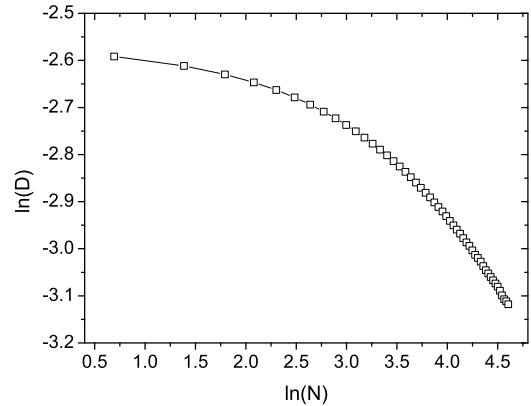}
 \caption{The quantity $\ln D$ versus $\ln N$ in the Dicke model, with parameters
 $\omega_0=\omega=1$, $\lambda_1=0.495$, and $\lambda_2=0.45$ ($\lambda_c=0.5$).}
 \label{eta0.1ln}
 \end{figure}

In the derivation of Eq.~(\ref{lpf}), we have used Eq.~(\ref{ee}),
which is obtained by diagonalizing an effective form of the 
exact Dicke Hamiltonian (\ref{DH}), given 
by Eq.~(\ref{Hhar}), with $n=2$ modes.
It is therefore important to assess the validity of the effective 
Hamiltonian in computing the fidelity $L_p$.
The effective Hamiltonian leads to an error inversely proportional to
 $N$, where $N$ is the number of atoms.
 For those quantities,
 for which there exist some contributions proportional to $N$, 
the effective Hamiltonian
 gives poor predictions \cite{res2005,vid2006,vid2007}.
However, for quantities like the fidelity $L_p$ of ground states, 
there is no such contribution. Hence,
the effective Hamiltonian is expected to correctly describe the behavior
of $L_p$ in the critical region.
To substantiate such expectation,
we have compared the prediction $L_p(\lambda_1,\lambda_2)$ of
 Eq.(\ref{lpf}) with $L_p^{N}(\lambda_1,\lambda_2)$, which is the corresponding
 fidelity numerically computed
 by direct diagonalization of the exact Hamiltonian (\ref{DH}) in a truncated Hilbert space.
 The truncated Hilbert space is obtained for a finite number $N$ of the atoms and by taking the lowest
 $N$ levels of the bosonic mode.
 We have studied the variation  of $D$ with $N$, where
 \be
 D=|L_p^N(\lambda_1,\lambda_2)-L_p(\lambda_1,\lambda_2)|.
 \ee
 With increasing $N$, the quantity $D$ exhibits a decay
faster than power law
 (see Fig.~\ref{eta0.1ln}) and slower than exponential.
 Therefore, it is reasonable to expect that the effective Dicke Hamiltonian 
provides the correct physical picture when
 computing the fidelity of ground states in the large $N$ limit.

\begin{figure}
 \includegraphics[width=\columnwidth]{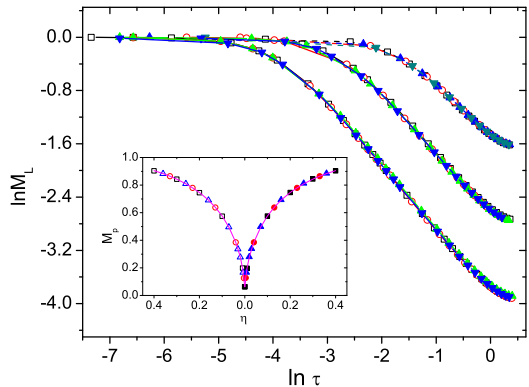}
 \caption{(Color online)
Dependence of the Loschmidt echo $M_L$ on
the rescaled time $\tau=\omega_1({\lambda_1})t$, for
 various values of $\lambda_1$ and $\lambda_2$,
 with different symbols representing different pairs ($\lambda_1,\lambda_2$).
 The three curves correspond, from top to bottom,
to $\eta = 10^{-2}$, $10^{-3}$, and $10^{-4}$.
An initial Gaussian decay is followed by $1/t$ decay,
 as predicted in Eq.~(\ref{Mt-sc-sa}).
 Inset: Dependence of the minimum value $M_p$ of $M_L$
on $\eta$ for various values
of $\lambda_1$ and $\lambda_2$ (open symbols stand for the normal phase and
solid ones for the super-radiant phase).
 The fitting curves are given by $M_p=2\sqrt{\eta}/(1+\eta ) $ in both phases.
 }
 \label{leteta}
 \end{figure}

Next, we discuss the LE.
For the Dicke model, it can be analytically proved
that the LE is an oscillating function of time with
period $T=\pi/\omega_1({\lambda_1})=T_1/2$.
This period diverges when
$\lambda_1$ approaches the critical point $\lambda_c$
and for times shorter than $T/2$ the LE decays
according to the above semiclassical prediction.
Indeed, numerical simulations in Fig.~\ref{leteta} show that
the LE is a function of $\eta$ and of the rescaled time
$\tau=\omega_1(\lambda_1)t$.
Moreover, in the super-radiant phase the LE decays
in the same manner as in the normal phase.
Finally, we have studied the minimum value of the LE,
denoted by $M_p$, as a function of $\lambda_1$
and $\lambda_2$.
Since this quantity is time-independent,
according to previous scaling arguments
it should be a function of the ratio $\eta$ only.
Such expectation is confirmed by our numerical
simulations (see the inset of Fig.~\ref{leteta}).

\section{Scaling for the LMG model}
\label{sec:LMG}

 In the two-orbital Lipkin-Meshkov-Glick (LMG) model for $N$ interacting particles,
 in terms of the total spin operator for its collective motion,
 $S_\alpha$ ($\alpha=x,y,z$), the Hamiltonian can be written as,
 \be
 H(\gamma,h) = -\frac{2}{N}(S^2_x+\gamma S^2_y) - 2hS_z + (1+\gamma)/2.
 \label{hlmg}
 \ee
 As shown in Ref.~\cite{DV05}, in the thermodynamic limit, making use of the Holstein-Primakoff
 transformation and of a standard Bogoliubov transformation,
the Hamiltonian can be diagonalized,
 \begin{eqnarray} \label{lmg}
 H(\gamma,h)& = & \Delta {a}_{\Theta}^{\dag}{a}_{\Theta},
 \end{eqnarray}
 where
\begin{eqnarray}
\Delta =2[(h-1)(h-\gamma)]^{1/2}, \  \
\tanh\Theta=\frac{1-\gamma}{2h-1-\gamma} \ \label{pa}
\end{eqnarray}
 for $h>1$,
\begin{eqnarray} \label{Dbrok}
\Delta =2[(1-h^2)(1-\gamma)]^{1/2}, \  \
\tanh\Theta=\frac{h^2-\gamma}{2-h^2-\gamma } \ \label{pb}
\end{eqnarray}
 for $h<1$, and ${a}_{\Theta}^{\dag}$ and ${a}_{\Theta}$ are bosonic creation and annihilation operators.

Equations (\ref{pa}) and (\ref{pb}) show that when $h$ approaches $1$ from both sides,
 $\Delta \to 0$.
 This implies that the system undergoes a quantum phase transition at the critical point $h_c=1$.
 The phase with $h>1$ is usually called the symmetric phase
 and the phase with $h<1$ the broken phase \cite{DV05}.

In the thermodynamic limit, the same scaling law as
in Eq. (\ref{lpf}) can be derived analytically in the vicinity
of the critical point. 
In fact, for a fixed $\gamma$ the ground state $|0\rangle_{\Theta_{2}}$
for $h=h_2$ has the following expansion
 on the basis $|n\rangle_{\Theta_{1}}$ of the eigenstates of
Hamiltonian (\ref{lmg}) at
$h=h_1$ \cite{WZW10}:
\begin{eqnarray} \label{0expan}
 |0\rangle_{\Theta_{2}}= \frac{1}{\sqrt{C}}\sum^{\infty}_{n=0}
 \sqrt{\frac{(2n-1)!!}{(2n)!!}}\tanh^{n}
\left(\frac{\Theta_{2}-\Theta_{1}}{2}\right)|2n\rangle_{\Theta_{1}},
 \end{eqnarray}
 where \textsl{C} is a normalization constant,
 \begin{eqnarray}
 C& = &\left [ 1-\tanh^{2} \left ( \frac{\Theta_{2}-\Theta_{1}}{2} \right ) \right ]^{-1/2}.
 \end{eqnarray}
Then, it is ready to find that
 \begin{eqnarray} \label{Lpt}
 L_{p}(h_{1},h_{2})& = &\left [ 1-\tanh^{2} \left ( \frac{\Theta_{2}-\Theta_{1}}{2}\right ) \right ]^{1/4}.
 \end{eqnarray}
In the vicinity of the critical point $h_c=1$,
 the right hand side of Eq.~(\ref{Lpt}) can be simplified further.
 In fact, in the symmetric phase, from Eq.~(\ref{pa}), one obtains, up to terms of higher order in
$h-h_c$,
 \begin{equation}
 \tanh\frac{\Theta}{2} 
 =1-2\left ( \frac{h-1}{1-\gamma}\right )^{1/2}.
\end{equation}
 This gives
 \begin{eqnarray}
 \tanh\left(\frac{\Theta_{2}-\Theta_{1}}{2}\right)=
\frac{\eta^{1/2}-1}{\eta^{1/2}+1},
\label{eq:tanhLMG}
 \end{eqnarray}
 where $\eta=(h_{1}-1)/(h_{2}-1)$.
After inserting (\ref{eq:tanhLMG}) into (\ref{Lpt}), we 
obtain the same expression (\ref{lpf}) for $L_p$ 
as for the Dicke model. 

For the LE, making use of analytical results in the
symmetric phase \cite{WZW10} and its generalization to the broken phase,
similar scaling behaviors as shown in Fig. \ref{leteta} have
also been found.

\section{Conclusions}
\label{sec:discussion}

To summarize, we have proved
a scaling  property for time-independent metric quantities such as the
fidelity and the participation ratio.
The scaling is valid for models like the Dicke model
and the LMG model,
whose QPT can be described in terms
of a single bosonic zero mode.
Moreover, also time-dependent quantities such as
the Loschmidt echo, exhibit the same
scaling provided time is measured in units of the inverse
frequency of the critical mode.

Our scaling arguments
showing $\eta$-dependence of static quantities can be generalized
to the cases of more than one zero mode, provided
appropriate new restrictions in the coefficients
$P_{ij}$ and $Q_{ij}$ in (\ref{c-sum}) are introduced.
On the other hand, the scaling for time-dependent
quantities cannot be extended in a straightforward way
to the case of more than one zero mode,
when the corresponding frequencies $\omega_i(\lambda)$ have different
scaling behaviors and, in contrast to the case of a single zero mode,
there is no natural rescaling of time by means of a single frequency.

Finally, we remark that in our theory we compute the fidelity
in the thermodynamic limit. The obtained results imply
that many-body systems whose QPT can be described in terms of a single
bosonic zero mode do not exhibit the Anderson orthogonality 
catastrophe~\cite{anderson}. That is to say, the ground states
corresponding to two nearby values of the controlling parameter
are not orthogonal at the thermodynamic limit, provided these
two values belong to the same phase. 

\acknowledgments
WW, PQ, and QW acknowledge support by the Natural Science
Foundation of China under Grant
Nos.~10775123 and 10975123, the National Fundamental Research
Programme of China Grant No.2007CB925200, and `Boshidian'
Foundation of the Ministry of Education of China.
GB and GC acknowledge support by
MIUR-PRIN 2008 and by Regione Lombardia.

\appendix
\section{Derivation of Eq.~(\ref{lpf})}
\label{sec:disckeanalytical}

In this section, we give a derivation of Eq.~(\ref{lpf})
in the main text for the fidelity
 of two ground states at $\lambda_1$ and $\lambda_2$.
 For this purpose, we use the following expression of the fidelity given in
Ref.~\cite{zan2006},
 \be
 L_p = \frac{2\left\{[\det A_{\lambda_2}]/[\det A_{\lambda_1}]\right\}^{1/4}}
 {[\det (1+A_{\lambda_1}^{-1}A_{\lambda_2})]^{1/2}},
 \label{lpa}
 \ee
 where $A_{\lambda}=U^{-1}M_{\lambda}U$, $M_{\lambda}=\mathrm{diag}
 [e_{1}^{\lambda},e_{2}^{\lambda}]$, and U is an orthogonal matrix,
 \be
 U=\left[
     \begin{array}{cc}
       c & -s \\
       s & c \\
     \end{array}
   \right] .
   \nonumber
 \ee
 Here, $c=\cos \gamma$, $s=\sin \gamma$, with
 \[ \gamma=\frac{1}{2}\arctan[4\lambda
 \sqrt{\omega\omega_0}/(\omega^2+\omega^2_0)]. \]
 It is straightforward to verify the following relations:
 \be
 \det A_{\lambda} = e_1^{\lambda}e_2^{\lambda},
 \ee
 \bey
 \det (1+A_{\lambda_1}^{-1}A_{\lambda_2}) &=& 1+\mathrm{Tr}(A_{\lambda_1}^{-1}A_{\lambda_2})\nonumber \\
 &+& [\det A_{\lambda_1}]^{-1}\det A_{\lambda_2},
 \label{tr}
 \eey
 and 
 \be
 \mathrm{Tr}(A_{\lambda_1}^{-1}A_{\lambda_2})=\frac{e_1^{\lambda_2}}{e_1^{\lambda_1}}+
 \frac{e_2^{\lambda_2}}{e_2^{\lambda_1}}.
 \ee

In the normal phase of the Dicke model, energies
$e_{1,2}^{\lambda}$ are given by Eq.~(\ref{ee}). 
In the neighborhood of the critical point $\lambda_c$,
from Eq.~(\ref{ee}) we get, up to terms of higher order in
$\lambda_c-\lambda$,
 \be
 e_{1}^{\lambda} = \Big[ \frac{8\lambda_{c}(\lambda_{c}-\lambda)\omega\omega_{0}}
 {\omega_{0}^{2}+ \omega^{2}} \Big] ^{1/2}.
 \label{edl}
\ee
 Then, we have $e_{1}^{\lambda_{2}}/e_{1}^{\lambda_1}= (1/\eta) ^{1/2} $.
 Using again Eq.~(\ref{edl}), we obtain
 \bey
 \frac{\det A_{\lambda_2}}{\det A_{\lambda_1}} &=& \sqrt{1/\eta},\label{dett}\\
 \mathrm{Tr}(A_{\lambda_1}^{-1}A_{\lambda_2}) &=& \sqrt{1/\eta}+1,
 \label{det}
 \eey
 where $e_2^{\lambda_2}/e_2^{\lambda_1}= 1$ has been used
in the vicinity of the critical point.
 Substituting the above results into Eq.~(\ref{lpa}), one finds
 Eq.~(\ref{lpf}).
 By the same method, the same expression of the fidelity 
can be obtained in the super-radiant phase.


\begin{thebibliography}{99}
\bibitem{footnote}
In previous studies of both the fidelity and the LE,  $\lambda_1$
and $\lambda_2$ were usually chosen to be sufficiently close to each other;
as a result, the dependence on
$(\lambda_1 - \lambda_c)$ and that on $(\lambda_2-\lambda_c)$ are similar.
\bibitem{foot-neta}
 The $\eta$-scaling to be discussed below does not exist, e.g., for Hermitian operators like
 $\hat{A}=\hat{c}^\dag_1(\lambda_1)\hat{c}^\dag_1(\lambda_1) +
\hat{c}_1(\lambda_1)\hat{c}_1(\lambda_1)
 +\hat{c}^\dag_1(\lambda_2)\hat{c}_1(\lambda_2)$ with
$\theta_c = \theta_c(\lambda_1)$
 a function of $\lambda_1$ only.
 This is not strange, because in the representation of
$\hat H(\lambda_2)$ it is seen clearly that
 this operator has separate dependence on
$\lambda_1$ and $\lambda_2$ by its definition.
\bibitem{foot-function relation}
The monotonicity is not strictly necessary and one can also
consider more complicated instances where
the relation
 $F(\Delta \lambda_1, \Delta \lambda_2)=d$
 implies a function relation
 between $\Delta \lambda_1$ and $\Delta \lambda_2$, written as
 $\Delta \lambda_2 = g(\Delta \lambda_1,d,\nu)$, where $\nu$ is an
 integer quantity independent of $\Delta \lambda_1$.
\end{thebibliography}
  \end{document}